# A Novel Approach for Estimating Largest Lyapunov Exponents in One-Dimensional Chaotic Time Series Using Machine Learning

Andrei Velichko*, Maksim Belyaev and Petr Boriskov

Institute of Physics and Technology, Petrozavodsk State University, 185910 Petrozavodsk, Russia

*Email: velichkogf@gmail.com

**Abstract**

Understanding and quantifying chaos from data remains challenging. We present a data-driven method for estimating the largest Lyapunov exponent (LLE) from one-dimensional chaotic time series using machine learning. A predictor is trained to produce out-of-sample, multi-horizon forecasts; the LLE is then inferred from the exponential growth of the geometrically averaged forecast error (GMAE) across the horizon, which serves as a proxy for trajectory divergence.

We validate the approach on four canonical 1D maps—logistic, sine, cubic, and Chebyshev—achieving $R^2_{pos} > 0.99$ against reference LLE curves with series as short as $M = 450$. Among baselines, KNN yields the closest fits (KNN-R comparable; RF larger deviations). By design the estimator targets positive exponents: in periodic/stable regimes it returns values indistinguishable from zero.

Noise robustness is assessed by adding zero-mean white measurement noise and summarizing performance versus the average SNR over parameter sweeps: accuracy saturates for $SNR_m \gtrsim 30$ dB and collapses below ≈27 dB, a conservative sensor-level benchmark.

The method is simple, computationally efficient, and model-agnostic, requiring only stationarity and the presence of a dominant positive exponent. It offers a practical route to LLE estimation in experimental settings where only scalar time-series measurements are available, with extensions to higher-dimensional and irregularly sampled data left for future work.

## Lead Paragraph

**Chaos means that tiny differences in a system's state can quickly grow into very different outcomes. The rate of this growth is set by the largest Lyapunov exponent (LLE). Many classic estimators need long, clean recordings and often break down on short or noisy data. We propose a simple, data-driven alternative. We train a standard machine-learning predictor on a one-dimensional time series and then read the LLE from how the predictor's forecast error grows with the prediction horizon (using a geometric average of errors). On four 1D maps—logistic, sine, cubic, and Chebyshev—the method matches reference exponents ($R^2_{pos} > 0.99$) even with only ~450 samples. It is robust to white measurement noise: accuracy stays high for average SNR above about 30 dB and drops sharply below ~27 dB. The estimator targets positive exponents and returns values near zero in stable regimes. Because it needs only measured data, it can be applied in settings from lab experiments to field telemetry where models are unknown.**

## I. Introduction

Chaos theory explores the behavior of nonlinear dynamical systems that, despite being deterministic, exhibit unpredictably complex patterns sensitive to initial conditions. Such sensitivity, popularly known as the "butterfly effect" [1,2] underscores that small variations in the starting state of a system can lead to significantly divergent outcomes over time. One of the key quantitative measures for characterizing this sensitivity and the chaotic nature of these systems is the Lyapunov exponent [3–5].

Lyapunov exponents measure the exponential rate of divergence or convergence of initially close trajectories in a dynamical system, thereby quantifying the degree of instability within the system. A positive Lyapunov exponent indicates chaos, signifying that trajectories diverge exponentially, resulting in unpredictable long-term behavior. Conversely, negative exponents denote convergence and stability, while zero exponents typically correspond to neutral, periodic, or quasiperiodic behavior [1,2].

The accurate calculation and interpretation of Lyapunov exponents is essential, as these values are directly connected to system predictability and stability. Specifically, they allow researchers and practitioners to evaluate how quickly small errors or uncertainties grow, providing critical insights into the reliability of predictions made by models describing complex systems. Systems characterized by predominantly positive Lyapunov exponents, such as those encountered in meteorology or ecological modeling, present significant challenges due to their reduced predictability and inherent sensitivity to small disturbances [6–8].

Accurate estimation of Lyapunov exponents drives progress in diverse fields. In meteorology and ecology, they enable real-time adaptive corrections that sharpen forecasts and optimize resource use [6]. In mechanical and aerospace engineering—e.g., helicopter flight dynamics—they expose incipient instabilities, helping engineers design safer control strategies [6]. Electrical-power operators rely on them to trace transient-stability margins and make rapid security assessments after disturbances [7,9]. Underlying these successes is the positive Lyapunov exponent, which quantifies the exponential divergence of nearby trajectories and thereby captures a system's sensitivity to initial conditions—the hallmark of chaos [10,11].

Quantitatively, the magnitude of the positive Lyapunov exponent directly correlates with the rate at which prediction errors or uncertainties in initial states amplify over time. Hence, knowledge of this value is critical for predicting the reliability of long-term forecasts and understanding the limits of predictability inherent in the system under study [12–14].

From a practical and methodological perspective, calculating the positive Lyapunov exponent is crucial for the development and validation of computational algorithms used in chaos analysis. Advanced methods, including traditional numerical techniques and modern machine learning algorithms, have been specifically designed and refined to efficiently and accurately estimate this exponent, underscoring its central role in the practical analysis of complex dynamical systems [15,16].

Traditionally, these exponents have been calculated using various approaches, such as local trajectory integration [17], nonparametric regression [18], pseudo-orbits [19], eigenvalue decomposition [20], time series analysis [10,21], and symplectic methods for Hamiltonian systems [22]. Each method has distinct advantages, such as accuracy, robustness to data changes, or reduced computational complexity.

In the context of machine learning approaches for estimating Lyapunov exponents, two primary directions can be distinguished. In the first case, the exponent values are directly approximated by a trained model from input time series data [6,18,23,24]. Supervised learning methods—including regression trees [6], multilayer perceptrons, convolutional neural networks (CNNs), and long short-term memory (LSTM) networks — have demonstrated accurate predictions of stable exponents, reasonable approximations for unstable exponents, and modest success with neutral exponents, especially in locally homogeneous attractors. Reservoir computing, leveraging a high-dimensional dynamical system to replicate chaotic attractors, has proven effective for complex dynamics such as those in the Lorenz and Kuramoto-Sivashinsky systems [25]. CNN-based deep learning has successfully handled hyperchaotic scenarios [24].

In the second case, the exponents are estimated indirectly, based on the analysis of the divergence between two nearby trajectories [15,19,21]. The latter approach is typically utilized to estimate only the largest Lyapunov exponent (LLE), which, however, is the most critical indicator for characterizing the chaotic behavior of nonlinear systems.

In addition to the above-mentioned methods, another indirect but insightful approach involves estimating the Lyapunov exponent through the entropy of a time series, as these parameters are inherently interconnected. Typically, higher entropy corresponds to a higher Lyapunov exponent, reflecting increased dynamical instability and unpredictability. However, this relationship provides only an approximate estimation and should be interpreted with caution. Recent studies by our group have extended these concepts by introducing modern entropy metrics, such as Neural Network Entropy (NNetEn)[26] and Logistic Neural Network (LogNNet)[27], specifically designed to quantify complexity within neural network contexts. Both NNetEn and LogNNet quantify complexity through nonlinear transformations of input data using chaotic or reservoir computing principles, and thus are closely related to Lyapunov exponents and traditional entropy measures. In practical contexts such as nonequilibrium statistical mechanics and transport phenomena, the difference between positive Lyapunov exponents and Kolmogorov–Sinai entropy per unit time is critical to understanding fractal dynamics and chaotic transport processes [28].

Machine learning methods notably reduce computational cost compared to traditional methods, which often involve intricate mathematical operations and large-scale data handling. Nevertheless, their accuracy is influenced by the system's characteristics and attractor homogeneity. Collectively, these ML techniques offer a promising, computationally efficient approach to enhance the understanding and prediction capabilities for chaotic dynamical systems, complementing existing classical methodologies. These methods offer substantial improvements in computational efficiency, and potentially enhanced accuracy and robustness, providing a non-intrusive means of analyzing chaotic systems without directly relying on intensive mathematical computations [3,6].

In one-dimensional settings the canonical objects of study are discrete-time maps on an interval—logistic, sine, cubic, Chebyshev, tent, and circle—which provide ground-truthable testbeds for theory, algorithms, and benchmarking [29–31]. Purely one-dimensional continuous-time flows cannot exhibit chaos; therefore, practical 1D analyses either use such discrete maps directly or rely on experimentally obtained return/peak-to-peak maps that reduce higher-dimensional dynamics to an effective one-dimensional iterate [29,31]. Estimation techniques tailored to 1D data range from classical nearest-neighbor and reconstructed-state-space methods (Wolf, Rosenstein) to finite-time/finite-size variants and data-driven or machine-learning approaches designed for short or noisy records [29,31,32]. Applications remain broad but the 1D perspective is central wherever intrinsic sampling or event-based observation naturally yields scalar sequences and return maps:

turbulence diagnostics and climate complexity in physics/geoscience [30,32,33], stability assessment of microgrids and nonlinear control architectures in engineering [34–36], and gait stability and epidemic dynamics in biomedicine and epidemiology [37,38]. Additional areas—including microfluidic mixing and general time-series complexity—use Lyapunov-based measures to quantify sensitive dependence and structural change in effectively one-dimensional observables [39,40]. Across these domains, reviews emphasize two persistent challenges—measurement noise and limited record length—motivating estimators that remain accurate under realistic conditions and that report uncertainty transparently[29,31].

In this study, we introduce a novel approach for estimating the largest Lyapunov exponent (LLE) in one-dimensional chaotic tme series using machine learning. Building on the two directions outlined above, we train ML predictors to produce out-of-sample, multi-horizon forecasts and then infer the LLE from the exponential growth of geometrically averaged forecast errors (GMAE), used as a proxy for trajectory divergence. The estimator is deliberately scoped to the largest exponent—aligned with practical predictability analysis—and is evaluated on four canonical 1D discrete maps (logistic, sine, cubic, Chebyshev) that afford controlled access to regular/chaotic regimes and admit analytical or well-established reference LLEs for validation. For discrete maps we set $\Delta t$=1 (per iteration), while the formulation naturally carries over to real data sampled at a physical interval.

To assess robustness, we perform a systematic study with additive white measurement noise across multiple SNR levels and quantify bias/variance in the estimated LLE. We further benchmark strong, interpretable ML baselines (KNN, KNN-based reservoir, Random Forest) to decouple model choice from intrinsic data information. Focusing on 1D maps—an experimentally relevant testbed via return-map and peak-to-peak reductions—also strengthens reproducibility, enabling dense parameter sweeps and precise comparisons against reference LLEs. Extensions to higher-dimensional invertible maps and continuous-time systems are outlined as future work.

Contributions:
(i) a GMAE-based LLE estimator unifying ML forecasting with divergence-rate inference;
(ii) a comprehensive evaluation on four canonical 1D discrete maps (logistic, sine, cubic, Chebyshev) with consistent ground-truth comparisons;
(iii) a robustness analysis under additive white measurement noise; and
(iv) practical guidelines (choice of horizons, window length, temporal splits) for stationary 1D chaotic time series.

## 2. Methodology

### 2.1 Generation of Chaotic Time Series

In this study, chaotic time series were generated using four discrete maps: logistic, sine, cubic, and Chebyshev maps. Depending on the control parameter $r$, these maps can produce both regular and chaotic behavior. The following recursive formulas define each discrete map along with the ranges of the control parameter $r$:

$$\begin{cases} x_{\mathrm{m}+1} = r \cdot x_{\mathrm{m}}\left(1 - x_{\mathrm{m}}\right), \; 3.5 \le r \le 4, \; \textit{Logistic map} \\ x_{\mathrm{m}+1} = r \cdot \sin(\pi x_{\mathrm{m}}), \; 0.85 \le r \le 1, \; \textit{Sine map} \\ x_{\mathrm{m}+1} = r \cdot x_{\mathrm{m}}\left(1 - x_{\mathrm{m}}^2\right), \; 2.3 \le r \le 3, \; \textit{Cubic map} \\ x_{\mathrm{m}+1} = \cos\left(r \cdot \arccos(x_{\mathrm{m}})\right), \; 1 \le r \le 10, \; \textit{Chebyshev map} \end{cases} \tag{1}$$

Each interval of the control parameter $r$ was divided into 500 equidistant points. For each of these points, a time series of 10,000 data points was generated, with the initial 1,000 points discarded to mitigate transient effects. Thus, for each chaotic map under consideration, we obtained 500 distinct time series corresponding to different values of the parameter $r$.

### 2.2 Standard Method for Calculation of Lyapunov Exponents for Discrete Maps

Lyapunov exponent $\lambda$ characterizes the regularity of a dynamic system, indicating how two trajectories in phase space, initially separated by distance $\delta_0$, diverge or converge over time $t$:

$$|\delta(t)| = |\delta_o| \cdot \exp(\lambda \cdot t). \tag{2}$$

For the discrete maps described in section 2.1 there is a single Lyapunov exponent. In general, however, a dynamical system admits a spectrum of Lyapunov exponents whose cardinality equals the phase-space dimension. Throughout this work we explicitly target the largest Lyapunov exponent (LLE, $\lambda_{max}$), which predominantly determines predictability. To avoid ambiguity with "positive" versus "maximum" exponents—and with the acronym used for maximum likelihood estimation—we do not use "MLE". Hereafter we denote the quantity of interest simply by $\lambda=\lambda_{max}$ .

An advantage of discrete maps is the straightforward theoretical estimation of the LLE. Considering the discrete nature of the time series, Eq. (2) implies the following standard expression for $\lambda$ [41]:

$$\lambda = \lim_{m \to \infty} \left( \frac{1}{m} \sum_{i=0}^{m-1} \ln\left( \left| f'(x_i) \right| \right) \right), \tag{3}$$

Here $f$, represents the recursive relationship defining each map (see equation (1)). For each of the 500 control parameter values $r$, the Lyapunov exponent was calculated using equation (3). The resulting curves served as reference values for evaluating the accuracy of the proposed method and computing error metrics (see section 2.5).

### 2.3 Calculation of the Largest Lyapunov Exponent Using Machine Learning

Based on Eq. (2) and discrete sampling with step $\Delta t$, the trajectory separation after $N$ steps satisfies

$$\ln(|\delta(N)|) = \ln(|\delta_0|) + \lambda\, \Delta t\, N. \tag{4}$$

In our approach, the out-of-sample forecast error of a machine-learning predictor is used as a proxy for the absolute separation $|\delta(N)|$. For each prediction horizon $N$ we define the absolute errors on the test set ($s$ samples) as

$$e_i(N) = \left| Y_i(N) - Y_i^{ml}(N) \right|, i = 1, \ldots, s, \tag{5}$$

where $Y_i(N)$ and $Y_i^{ml}(N)$ are the true and predicted values, respectively.

To aggregate errors consistently with the "mean of logarithms" principle underlying Eq.(3) we use the geometric mean absolute error (GMAE):

$$\mathrm{GMAE}(N) = \left(\prod_{i=1}^{s} e_i(N)\right)^{1/s}. \tag{6}$$

Empirically, GMAE($N$) grows approximately exponentially with $N$. We therefore fit

$$\mathrm{GMAE}(N) \approx A \exp(\lambda_{ml} \Delta t N), \tag{7}$$

which is equivalent to a linear model for the log-transformed curve,

$$\ln \mathrm{GMAE}(N) = \ln A + (\lambda \Delta t) N. \tag{8}$$

Let $b$ be the slope of the linear approximation in Eq. (8) over the range of $N$ where the growth is quasi-exponential. The estimate of the largest Lyapunov exponent is then

$$\lambda_{ml} = \frac{b}{\Delta t}. \tag{9}$$

Procedure: (i) Split the data into train/test subsets for each horizon $N$ (respecting temporal order). (ii) Train the predictor on the training subset. (iii) Compute test errors $e_i(N)$ and GMAE($N$) by Eq. (6). (iv) Regress ln(GMAE($N$)) against $N$ and obtain $b$ (Eq. (8)).

Normalization and units: For discrete maps, one iteration is taken as a unit time step, $\Delta t = 1$. Equations (7)–(9) then reduce to $ln\ \mathrm{GMAE}(N) \approx \ln A + \lambda_{ml} N$ and $\lambda_{ml} = b$.

Rationale and scope: ML predictors approximate the training-set trajectory manifold; thus the out-of-sample forecast error quantifies the average divergence of test trajectories from their closest learned counterparts. The estimator targets $\lambda_{max} > 0$, in non-chaotic regimes ($\lambda \le 0$) the slope tends to zero and no positive exponent is reported.

To estimate the maximum Lyapunov exponent using machine learning methods ($\lambda_{ml}$), the following procedure is implemented:

1. For each prediction horizon $N = 1..K$, the original time series $X = \{x_1, x_2, \ldots x_M\}$ of length $M$ is divided into segments of length $L$ (where $L < M$), creating the input feature set $X_{ml}$ as follows: $X_{ml} = \{\{x_1, x_2, \ldots x_L\}, \{x_2, x_3, \ldots x_{L+1}\}, \ldots \{x_{M-L-N-1}, x_{M-L-N}, \ldots x_{M-N}\}\}$. Correspondingly, the target variable set $Y_{ml}$ is formed by selecting the points located exactly $N$ steps ahead of each segment: $Y_{ml} = \{x_{L+N}, x_{L+N+1}, \ldots, x_M\}$. Thus, the input segments ($X_{ml}$) represent the features, and the future values ($Y_{ml}$) represent the target variable the algorithm aims to predict.
2. Both feature set $X_{ml}$ and target variable set $Y_{ml}$ are split into training and testing subsets, maintaining the same test subset proportion $P$ for each prediction horizon $N$. The training subsets are used to train the machine learning algorithms, while the testing subsets are used to evaluate the GMAE($N$).
3. Compute absolute test errors $e_i(N)$ according to Eq. (5) and aggregate them by the geometric mean, obtaining GMAE($N$) per Eq. (6).
4. Fit a straight line to ln GMAE($N$) versus $N$ as in Eq. (8); the slope b gives the estimate of the largest Lyapunov exponent.

In Eq. (5) we compare the true future $Y_i(N)$ with the forecast $Y_i^{ml}(N)$ produced from the same observed initial segment. Although the starting state is identical, the learned mapping (a surrogate of the $N$-step flow) inevitably incurs a small out-of-sample error at the first step. This error acts as an effective initial separation $\delta_0$. For chaotic dynamics, Eqs. (2)–(4) imply that the separation evolves as $\delta(N)| \approx |\delta 0| e^{\lambda N}$; hence the forecast error grows approximately exponentially with the horizon $N$. Aggregating errors by the geometric mean (Eqs. (6)–(8))—i.e., averaging log-errors—provides a consistent proxy for this growth and yields the slope $b$ used in Eq. (9) to estimate the largest Lyapunov exponent. In non-chaotic regimes ($\lambda \le 0$) the growth vanishes and the slope approaches zero, due to the fact that the forecast error does not increase with increasing $N$ for regular time series.

This interpretation relies on stationarity and local smoothness so that the first-step model error is small and the log-error growth is approximately linear over a range of horizons. We therefore report $\lambda_{ml}$ only on the linear portion of ln GMAE($N$) versus $N$ and treat the method as an estimator of largest $\lambda > 0$ rather than of the full Lyapunov spectrum.

Parameters of the proposed algorithm include the number of points $K$, segment length $L$, relative size of the test subset (to full dataset) $P$, and machine learning hyperparameters (see section 2.4). This method applies to stationary chaotic time series since the training-test split approach cannot adequately assess transient processes with changing amplitude values.

Specifically, the parameter $K$ represents the number of prediction horizons $N$ used in constructing the linear approximation (see Eq.6). The approximation starts from $N = 1$, and includes $K$ consecutive values: for example, if $K = 2$, the linear fit is performed using $N = 1, 2$; if $K = 4$, then $N = 1, 2, 3, 4$, and so on. It is evident that $K > 1$ must hold, since a minimum of two points is required to construct a linear approximation.

The parameter $L$ denotes the number of historical data points in each input segment used by the machine learning model for predicting future values.

### 2.4 Machine Learning Methods

To estimate $\lambda_{ml}$ using MAE, several machine learning methods were selected, including both standard algorithms — such as k-nearest neighbors (KNN) [42], random forest (RF) and a novel reservoir-based method employing KNN (R-KNN). The proposed R-KNN approach is described in detail in subsection 2.4.1.

Hyperparameters for KNN included the number of neighbors, weighting type, and distance function. For RF, hyperparameters included the number of trees, split quality evaluation criterion, maximum tree depth, minimum split size, and minimum leaf size.

#### 2.4.1 Reservoir-based Network with KNN (R-KNN)

Reservoir networks consist of interconnected neurons whose internal connection weights and input weights (used to pass features into reservoir neurons) are randomly initialized and remain unchanged during training. Only the weights connecting the reservoir neurons to the output linear layer are adjusted throughout training. This architecture represents a type of reservoir neural network known for high-accuracy predictions of chaotic time series[43]. Prominent examples of reservoir-based networks include echo state networks (ESN)[44], LogNNet[45], and liquid state machines (LSM)[43].

In this study, KNN was proposed as the output layer method, predicting time series values by comparing reservoir neuron activities. Given that the reservoir typically contains numerous neurons, principal component analysis (PCA)[46] is employed to reduce the dimensionality of the feature vector fed into KNN, significantly simplifying data representation with minimal loss of information. Thus, reservoir neuron outputs undergo PCA transformation, after which KNN predicts the time series values.

The hyperparameters for the proposed method include reservoir properties (number of neurons, connectivity, spectral radius, activation function), input layer parameters (scaling factor, connectivity with reservoir neurons), PCA parameters (number of components), and the aforementioned KNN hyperparameters.

### 2.5 Method for Optimizing Parameters

To evaluate the accuracy of the calculated $\lambda_{\mathrm{ml}}$, the deviation of $\lambda_{\mathrm{ml}}(r)$ from the reference values $\lambda(r)$ obtained using the standard calculation method (section 2.2) was assessed. The coefficient of determination $R^2$ was chosen as the metric for evaluating calculation errors, computed as follows. First, the average value $\lambda'$ was calculated for all $T = 500$ points in the range of the parameter $r$:

$$\lambda' = \frac{1}{T}\sum_{i=1}^{T}\lambda_i \ . \tag{10}$$

Next, the residual sum of squares $SS_{\mathrm{res}}$ and total sum of squares $SS_{\mathrm{tot}}$ were computed:

$$SS_{\mathrm{res}} = \frac{1}{T}\sum_{i=1}^{T}\left(\lambda_i - \lambda_{\mathrm{ml}_i}\right)^2 , \tag{11}$$

$$SS_{\mathrm{tot}} = \sum_{i=1}^{N}\left(\lambda_i - \lambda'\right)^2 . \tag{12}$$

Using these sums, the coefficient of determination $R^2$ was calculated:

$$R^2 = 1 - \frac{SS_{\mathrm{res}}}{SS_{\mathrm{tot}}} \ . \tag{13}$$

An $R^2$ value closer to 1 indicates a stronger agreement between $\lambda(r)$ and $\lambda_{\mathrm{ml}}(r)$. It should be noted that the proposed method yields only positive values of $\lambda_{\mathrm{ml}}$. Therefore, error evaluation was performed both for the entire curve ($R^2_{\mathrm{tot}}$) and exclusively for points where $\lambda(r) \geq 0$ ($R^2_{\mathrm{pos}}$).

For each machine learning method and discrete map, parameters described in sections 2.3 and 2.4 were optimized to maximize $R^2_{\mathrm{pos}}$, which is equivalent to maximizing $R^2_{\mathrm{tot}}$.

## 3. Experimental Results

### 3.1 Analysis of the Proposed Method Efficiency

Table 1 presents the values of $R^2_{\mathrm{pos}}$ and $R^2_{\mathrm{tot}}$ for various machine learning methods and discrete maps. It should be emphasized that the presented results correspond to the optimal parameters that minimize the error.

Table 1. Error metrics ($R^2_{\mathrm{pos}}$ and $R^2_{\mathrm{tot}}$) for $\lambda_{\mathrm{ml}}(r)$ relative to reference $\lambda(r)$ for various machine learning methods.

| **ML Algorithm** | **Logistic Map** | | **Sine Map** | | **Cubic Map** | | **Chebyshev Map** | |
|---|---|---|---|---|---|---|---|---|
| | $R^2_{\mathrm{pos}}$ | $R^2_{\mathrm{tot}}$ | $R^2_{\mathrm{pos}}$ | $R^2_{\mathrm{tot}}$ | $R^2_{\mathrm{pos}}$ | $R^2_{\mathrm{tot}}$ | $R^2_{\mathrm{pos}}$ | $R^2_{\mathrm{tot}}$ |
| KNN | 0.998 | 0.796 | 0.997 | 0.853 | 0.999 | 0.919 | 0.999 | 0.999 |
| RF | 0.997 | 0.796 | 0.997 | 0.850 | 0.997 | 0.918 | 0.997 | 0.997 |
| KNN-R | 0.998 | 0.796 | 0.996 | 0.852 | 0.999 | 0.919 | 0.999 | 0.999 |

The metric $R^2_{\mathrm{pos}}$ for all maps exceeds 0.99, indicating a good degree of correspondence between $\lambda_{\mathrm{ml}}(r)$ and the reference curves $\lambda(r)$. The best match was observed for the Chebyshev map, achieving near-perfect results ($R^2_{\mathrm{pos}} = 0.999$). The highest accuracy was achieved by the KNN model for the cubic and Chebyshev maps and the RF model for logistic and sine maps. The performance of the KNN-R model closely mirrored that of the standard KNN.

It is important to note that the $R^2_{\mathrm{tot}}$ metric is generally less meaningful because the proposed methodology cannot estimate negative Lyapunov exponents from slope angles. Additionally, during time series generation, initial transient values were discarded, resulting in constant or repeating values for segments corresponding to negative Lyapunov exponents. Therefore, $R^2_{tot}$ is presented solely alongside $R^2_{pos}$ for completeness.

Figure 1 plots $\lambda_{ml}(r)$ obtained with the KNN estimator for all four maps alongside the reference lambda(r). In chaotic regions ($\lambda(r) > 0$) the curves align closely. Near periodic windows, the estimates drop toward zero, consistent with Sec. 2.3.

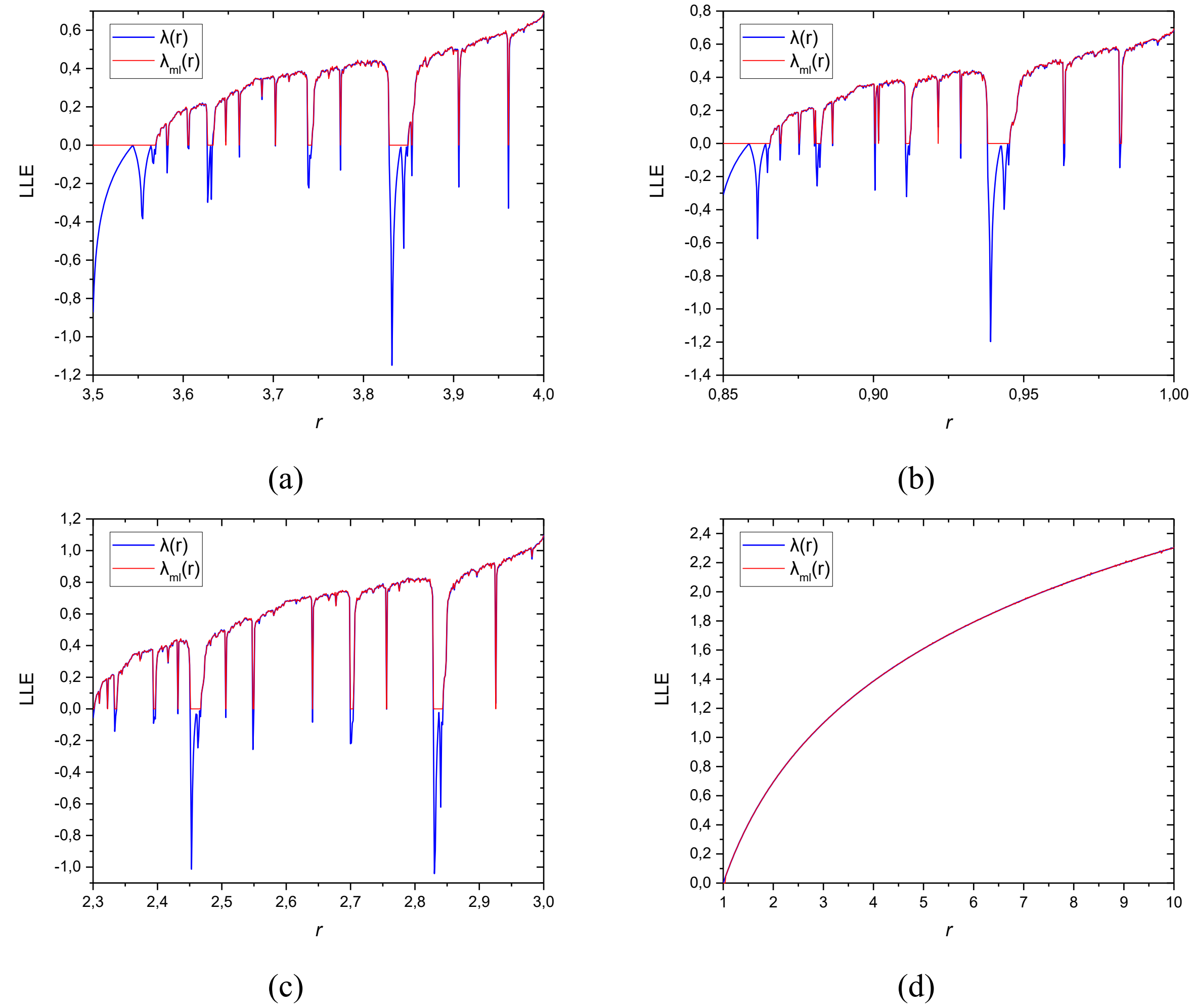

Figure 1. Estimated LLE curves $\lambda_{ml}(r)$ using KNN for (a) logistic, (b) sine, (c) cubic, and (d) Chebyshev maps, overlaid with the reference lambda(r). In parameter ranges where $\lambda(r) < 0$ (periodic windows), $\lambda_{ml}(r)$ tends to approximately zero by design.

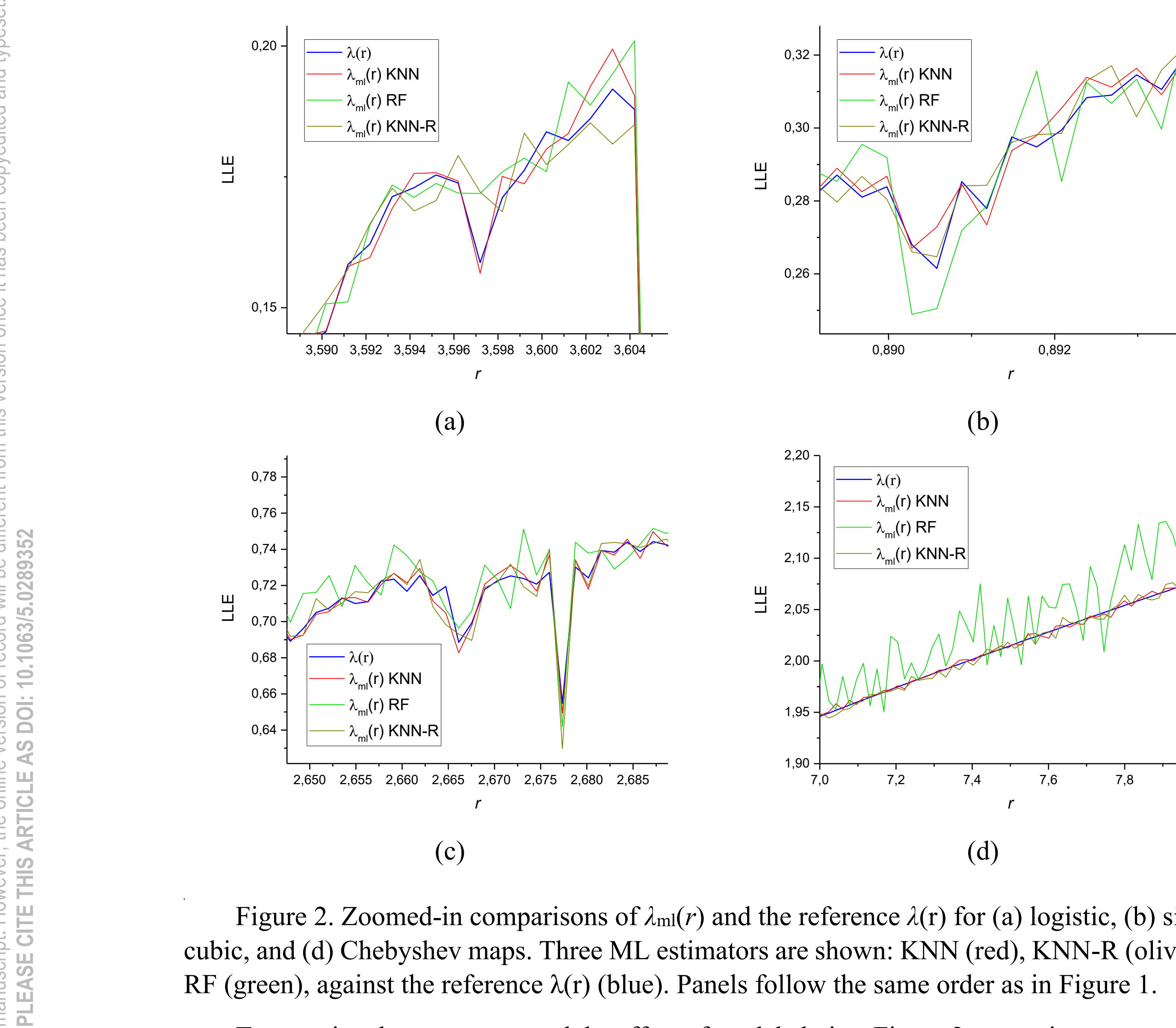

Figure 2. Zoomed-in comparisons of $\lambda_{\mathrm{ml}}(r)$ and the reference $\lambda$(r) for (a) logistic, (b) sine, (c) cubic, and (d) Chebyshev maps. Three ML estimators are shown: KNN (red), KNN-R (olive), and RF (green), against the reference λ(r) (blue). Panels follow the same order as in Figure 1.

To examine the agreement and the effect of model choice, Figure 2 zooms into representative r-intervals (same order as Fig. 1) and overlays KNN, KNN-R, and RF. In all panels, KNN tracks the reference λ(r) most closely; KNN-R yields a slightly smoother trace with comparable bias, while RF shows the largest deviations—most prominently for the Chebyshev map in panel (d), where RF exhibits high-frequency fluctuations and over/under-shoots relative to λ(r). The same trend is visible for the logistic and sine maps (panels (a)–(b)), with the logistic case showing the clearest gap between KNN and RF.

### 3.2 Influence of Parameters on Estimation Accuracy

To determine the optimal parameter values for calculating the LLE (see Section 2.3), we analyzed how the accuracy metric $R^2_{\mathrm{pos}}$ depends on the parameters *K*, *L*, and *P* for different machine learning models. Figure 3 illustrates the corresponding dependencies of $R^2_{\mathrm{pos}}$ for the logistic map.

(a)

(b)

(c)

**Figure 3.** Dependencies of prediction accuracy ($R^2_{pos}$) on segment length ($L$), number of points for slope estimation ($K$), and test subset ratio ($P$), illustrated for different machine learning algorithms (KNN, RF, KNN-R) on the logistic map.

At smaller values of $L$ (ranging from 1 to 4), the accuracy $R^2_{pos}$ exhibits weak dependency on $L$ for KNN and KNN-R. However, with further increases in segment length ($L > 4$), these algorithms demonstrate a pronounced reduction in estimation accuracy. For RF, a constant quasi-linear decrease in accuracy is observed starting from $L = 1$. The decrease in accuracy is quasi-linear for all curves on the specified intervals of parameters $L$ (except for $L = 9$ for KNN). Additional series elements, which diverge significantly and thus lack useful predictive information, contribute to decreased prediction accuracy. This deterioration, in turn, can lead to incorrect estimation of the LLE.

It should also be noted that when analyzing less chaotic time series, increasing $L$ might actually improve the accuracy of LLE estimation. However, for practical applications involving one-dimensional chaotic maps, selecting a minimal segment length $L = 1$ is sufficient and typically optimal.

Figure 3b illustrates the dependency of the prediction accuracy metric $R^2_{pos}$ on the number of points $K$, used to approximate the slope of the line ln GMAE($N$) in our proposed method. For all used ML algorithms, a decrease in accuracy is true when a certain value K is exceeded. This phenomenon arises because, at higher values of the prediction horizon $N$, the GMAE($N$) exceeds the standard deviation of the data itself, approaching a constant level regardless of further increases in $N$. Consequently, the slope of the line ln(GMAE) becomes artificially flattened when $K$ surpasses its optimal value, resulting in reduced accuracy. For RF algorithm, initially, as the number of points $K$ increases, the value of $R^2_{pos}$ rises, reflecting a more accurate estimation of the slope of the ln(GMAE) curve as a function of the prediction horizon $N$.

Based on the experimental results, optimal values of $K$ range from 4 to 7 points. Among the tested algorithms, KNN and KNN-R models exhibit the weakest dependence on this parameter, showing greater robustness to changes in $K$. Moreover, for time series with smaller LLE values, increasing $K$ beyond the established optimal range may further improve the estimation accuracy.

The dependency of the prediction accuracy metric $R^2_{pos}$ on the proportion $P$ (ratio of the test subset size to the entire dataset) is represented by a gradually declining curve (see Fig. 3c). For all tested algorithms, accuracy remains relatively stable within the range $0.2 \leq P \leq 0.8$, indicating robustness against moderate changes in the training-test split ratio. Only at extreme values, when the training subset is too small ($P = 0.9$) or the test subset is too small ($P = 0.1$), the accuracy $R^2_{pos}$ begin to noticeably decrease. This reduction in performance is attributed to insufficient data available for training ($P = 0.9$) or the small and therefore not very reliable for assessing accuracy, test subset ($P = 0.1$).

### 3.3 Influence of Time Series Length on Estimation Accuracy

An important characteristic of the proposed LLE estimation method is the minimal required time series length for which the LLE can still be estimated with sufficient accuracy. To investigate this, we constructed the $R^2_{pos}(M)$ curves for the KNN, RF, and KNN-R algorithms using the logistic map as a test case. For each series length $M$, the parameters were individually optimized to achieve the highest possible $R^2_{pos}$.

It is important to note that the reference curves $\lambda(r)$ were calculated using a series length of $M = 10{,}000$. The results are presented in Figure 4.

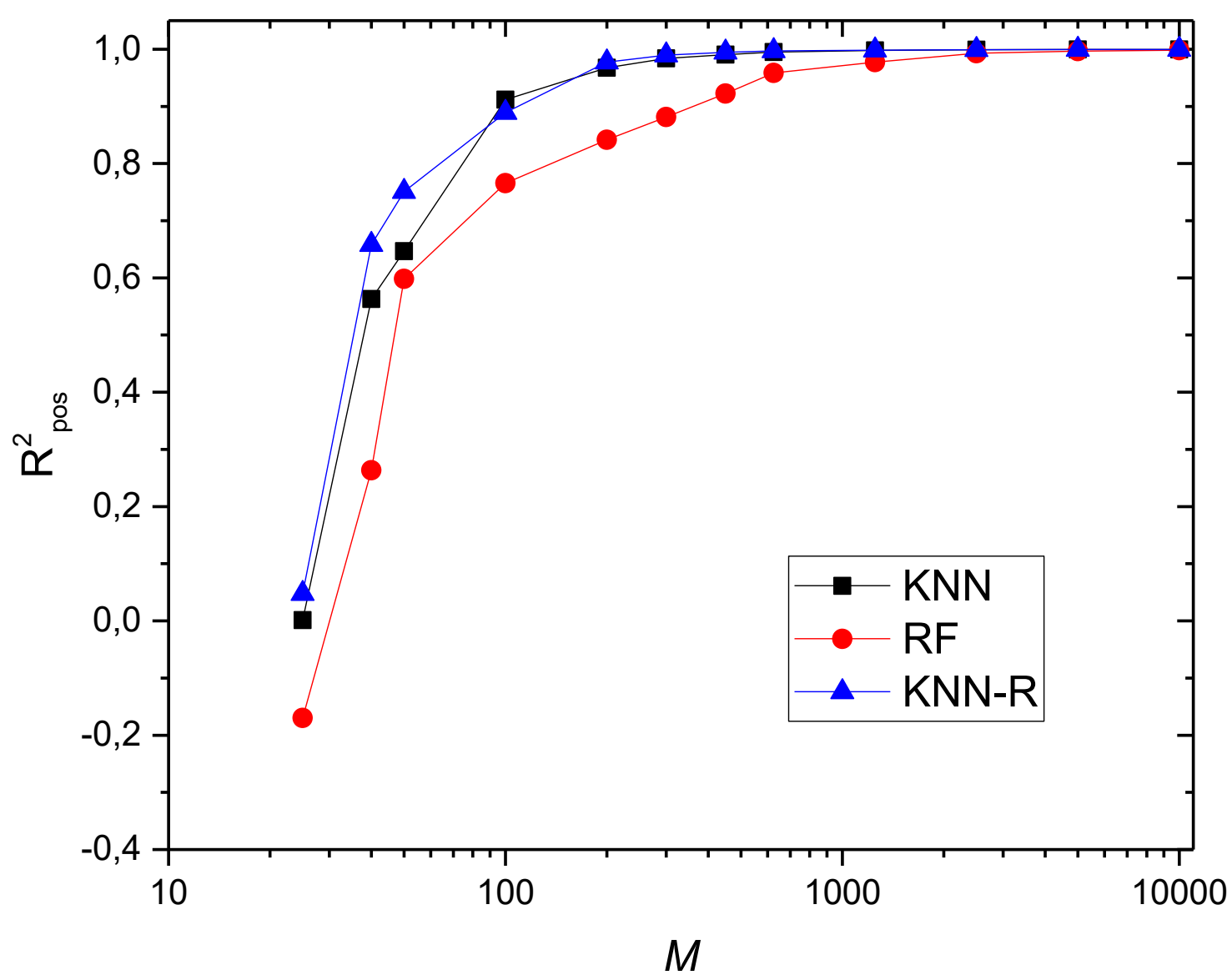

**Figure 4.** Dependence of $R^2_{pos}$ on time series length $M$ for different machine learning algorithms (KNN, RF, KNN-R) using the logistic map.

The findings indicate that for time series longer than $M = 200$ elements, the estimation accuracy for KNN and KNN-R exceeds $R^2_{pos} > 0.96$ and is only slightly (less than 5%) lower than the accuracy achieved with a time series of length 10,000. For the RF, such accuracy is achieved only at $M = 625$.

Among the tested models, the reservoir-based KNN-R algorithm demonstrated the best performance for shorter series: even with $M = 50$ elements, it achieved an accuracy of $R^2_{pos} = 0.751$, compared to $R^2_{pos} = 0.598$ for RF and $R^2_{pos} = 0.647$ for KNN. For time series shorter than 25 elements, the estimation accuracy was unsatisfactory for all methods, with $R^2_{pos}$ is near or below 0.

**3.4 Noise Robustness (additive white measurement noise)**

To evaluate robustness to measurement noise — a common condition in real-world applications — we added zero-mean white noise $x^n$ to the original time series $x$ (Sec. 2.1) and computed the LLE from the noisy observations while comparing against the clean reference $\lambda(r)$ (Eq. 3). The noise sequence $x^n$ had the same length as the map output ($M = 10{,}000$) and a fixed standard deviation common to all parameter values $r$. The signal-to-noise ratio for each series was defined by

$$\mathrm{SNR} = 10 \log \frac{\sum_{i=1}^{M} (x_i)^2}{\sum_{i=1}^{M} (x_i^n)^2} \,. \qquad (14)$$

Because the signal power depends on $r$, the SNR varies slightly across the 500 curves; therefore, we report the average SNR over $r$, denoted $\mathrm{SNR_m}$. Figure 5 shows the dependence of $R^2_{pos}$ (Sec. 2.5) on $\mathrm{SNR_m}$ for the logistic map using the KNN estimator (the most stable baseline in Fig. 2). Performance is high and saturates for $\mathrm{SNR_m} > 30$ dB ($R^2_{pos} \approx 0.90$ at 30 dB and $\geq 0.95$ for 40–90 dB). As $\mathrm{SNR_m}$ drops below 27 dB, accuracy declines rapidly; at $\mathrm{SNR_m} \approx 20$ dB the $R^2_{pos}$ falls near 0.1, and at still lower SNR it becomes negative, indicating that the noisy-data estimates no longer track the clean reference curve. This behavior is expected: we compare to $\lambda(r)$ of the noise-free map (Eq. (3)), while additive noise degrades forecasts and flattens the growth of errors. No denoising or noise-aware modeling was applied; thus the plot reflects a conservative, “sensor-level” robustness.

**Figure 5.** Noise robustness for the logistic map: $R^2_{pos}$ versus average $SNR_m$ using the KNN estimator.

### 3.5 Comparative analysis of calculation time

To compare the computational cost of the proposed method across different machine learning models, we conducted a performance benchmark using a time series of length $M$ = 10,000. For a more reliable estimation, each calculation was repeated 100 times, and the average computation $t_{comp}$ was recorded.

The calculations were performed using the python programming language (version 3.11.7) on the Windows 11 operating system, on a computer with an Intel Core i7 12700H processor and 32 GB of RAM without using a graphics processor.

The measurements were performed using the following parameters: $L$ = 2, $K$ = 4, and $P$ = 0.2. The results are summarized in Table 2.

**Table 2.** Average computation time per time series of length $M$ = 10,000 for different machine learning algorithms.

| Machine Learning Algorithm | Average Computation Time $t_{comp}$, ms |
|---|---|
| KNN | 101 |
| RF | 235 |
| KNN-R | 382 |

The fastest algorithm was KNN. The RF and KNN-R methods were approximately 2 to 4 times slower.

### 4. Conclusion

This study introduces and systematically validates a novel approach for estimating the largest Lyapunov exponent (LLE) in one-dimensional chaotic time series using machine learning. Rather than predicting exponents directly, we train ML predictors to produce out-of-sample, multi-horizon forecasts and infer the LLE from the exponential growth of the geometrically averaged forecast error (GMAE) as a function of the prediction horizon. The estimator is tailored to the largest exponent and is evaluated on four canonical 1D discrete maps (logistic, sine, cubic, Chebyshev) with established reference values.

Across maps, the method recovers the positive LLE with high correspondence to ground truth (for clean data, $R^2_{pos}$ >0.99), while requiring only moderate series length (e.g., $M > 450$). Among the tested baselines, KNN consistently yields the closest fits; KNN-R is comparable and smoother; RF exhibits larger deviations in some regimes (notably for Chebyshev). By construction the estimator targets $\lambda$max>0: in periodic or stable regimes it returns values indistinguishable from zero, providing a simple diagnostic of transitions to and from chaos.

Noise robustness. We injected zero-mean additive white measurement noise and quantified performance versus the average SNR over parameter sweeps. For the logistic map with KNN, accuracy saturates for $SNR_m \gtrsim 30$ dB ($R^2_{pos} \approx 0.9$–0.99) and collapses below ≈27 dB, where estimates no longer track the clean reference. This conservative "sensor-level" protocol used constant-amplitude noise across all parameter values; averaging over 500 points produced a stable global trend. Similar monotone behavior was observed for the other maps, with KNN most resilient at low SNR.

Applicability. Because the estimator relies only on learned input–output mappings and multi-horizon error growth, it can be applied to real time series when a single dominant positive exponent governs predictability. For discrete maps we take $\Delta t$=1; for real data the sampling interval sets the time units of the LLE.

Scope and limitations. All experiments used uniformly sampled 1D maps and the method estimates only positive LLE; negative exponents are reported as ~0. Accuracy degrades for very short records (e.g., $M < 50$) and depends on hyperparameters (window length $L$, horizons $K$, test ratio $P$) and model choice.

Future work. We plan to (i) extend the framework to higher-dimensional invertible maps and continuous-time systems, (ii) incorporate noise-aware modeling and denoising to improve low-SNR performance, and (iii) adapt the pipeline to nonstationary or transient dynamics via adaptive windowing and model selection.

**Acknowledgments**

Special thanks to the editors of the journal and to the anonymous reviewers for their constructive criticism and improvement suggestions.

This research was supported by the Russian Science Foundation (grant no. 22-11-00055-P, https://rscf.ru/en/project/22-11-00055/, accessed on 10 June 2025).

**Author Contributions**

Conceptualization, A.V.; Methodology, A.V., M.B.; Software, A.V. and M.B.; Validation, A.V. and P.B.; Formal analysis, P.B.; Investigation, A.V. and M.B.; Resources, A.V.; Data curation, A.V.; Writing — original draft preparation, A.V., M.B. and P.B.; Writing — review and editing, A.V., M.B. and P.B.; Visualization, M.B.; Supervision, A.V.; Project administration, A.V.; Funding acquisition, A.V. All authors have read and agreed to the published version of the manuscript.

**Conflicts of Interest**

The authors declare no conflicts of interest.

**Data availability statement**

The data that support the findings of this study are available from the corresponding author, upon reasonable request.